# Molecular Isomer Identification of Titan's Tholins Organic Aerosols by Photoelectron/Photoion Coincidence Spectroscopy Coupled to VUV Synchrotron Radiation.

*Barbara Cunha de Miranda[1,†], Gustavo A. Garcia[1], François Gaie-Levrel[1,2], Ahmed Mahjoub[3], Thomas Gautier[3,4], Benjamin Fleury[3], Laurent Nahon[1], Pascal Pernot[5], Nathalie Carrasco[3,6,*]*

[1] Synchrotron SOLEIL, DESIRS beamline, L'Orme des Merisiers, St Aubin 91192 Gif-sur-Yvette Cedex, (France).

[2] Laboratoire National de Métrologie et d'Essais – LNE (National Metrology Institute and Testing Laboratory), Chemistry and Biology Division, Gas and Aerosol metrology department, 1 rue Gaston Boissier, 75724 Paris Cedex 15 (France).

[3] Université Versailles St-Quentin ; Sorbonne Universités, UPMC Univ. Paris 06 ; CNRS/INSU, LATMOS-IPSL, 11 Boulevard d'Alembert, 78280 Guyancourt (France).

[4] NASA Postdoctoral Program, GSFC, Greenbelt, Maryland 20771 (USA).

[5] Laboratoire de Chimie Physique, UMR8000 CNRS/Université Paris-Sud, 91405 Orsay Cedex (France).

[6] Institut Universitaire de France, 103 Bvd St-Michel, 75005 Paris (France).

**ABSTRACT:** The chemical composition of Titan organic haze is poorly known. To address this issue, laboratory analogs named tholins are synthesized, and analyzed by methods requiring often an extraction process in a carrier solvent. These methods exclude the analysis of the insoluble tholins fraction and assume a hypothetical chemical equivalence between soluble and insoluble fractions. In this work, we present a powerful complementary analysis method recently developed on the DESIRS VUV synchrotron beamline at SOLEIL. It involves a soft pyrolysis of tholins at ~230°C and an electron/ion coincidence analysis of the emitted volatiles compounds photoionized by the tunable synchrotron radiation. By comparison with reference photoelectron spectra (PES), the spectral information collected on the detected molecules yields their isomeric structure. The method is more readily applied to light species (m/z ≤ 69), while for heavier ones the number of possibilities and the lack of PES reference spectra in the literature limit its analysis. A notable pattern in the analyzed tholins is the presence of species containing adjacent doubly-bonded N atoms, which might be a signature of heterogeneous incorporation of $N_2$ in tholins.

**Introduction**

Titan, the largest moon of Saturn, has a dense atmosphere mainly composed of nitrogen (98%), methane, molecular hydrogen, traces of hydrocarbons and nitrogenous compounds. Atmospheric photochemistry leads to the production of microscopic particles forming the brownish haze that permanently surrounds Titan [1]. The ongoing Cassini-Huygens space mission provided first insights on the chemical composition of these atmospheric organic aerosols. Mid- and Far-IR spectroscopy confirmed some aliphatic signatures [2-5], while the Aerosol Collector Pyrolyser experiment highlighted the presence of nitrogen in the refractory nucleus [6].

To complete this chemical overview, the knowledge about Titan's aerosols is improved by the study of analogous materials produced in the laboratory ("Titan's tholins"). The word "tholins" has been proposed in 1979 by Sagan and Khare [7] to name laboratory analogues of solid planetary aerosols. It comes from the Ancient Greek *thólos* and refers to the brownish color of the material. These are most often synthesized by maintaining a continuous plasma discharge in gaseous $N_2$-$CH_4$ mixtures [7-10], such as the PAMPRE reactor in LATMOS laboratory. This original setup is designed for atmospheric simulation, as the production and growth of the tholins occur directly in levitation inside the confined reactive plasma, avoiding any contact with a solid support during the growth process.

High-resolution mass spectrometry analysis has been performed in order to understand the $C_xH_yN_z$ chemical structure [11-16] of Titan's tholins. It was found to contain a complex mixture of molecules: mass spectra were actually obtained with more than one species per nominal mass at every mass between m/z 50 and 800. Van Krevelen representations enabled to identify their molecular formula, polymeric patterns and to decompose the complex organic matter into only about ten polymeric families [11-12]. But high resolution mass spectrometry does not solve isomeric ambiguity and cannot provide the chemical structure of the molecules. Moreover, the extraction method relies on the solubility of the material, which has been found to be mostly insoluble, even in a polar solvent: only 20-30% can be dissolved in methanol, and less than 1% in toluene [17]. The material analysis has therefore to be completed by other extractions techniques, such as pyrolysis.

Previous studies involving Titan's tholins pyrolysis were performed [18-20]. In [18] tholins were pyrolysed at four temperatures, 250, 400, 600 and 900°C, and analyzed by GC-GC-MS. Surprisingly, at 250°C, very few species were detected, among them acetonitrile was the major one. Because of this limitation, the authors focused on identifying the volatiles compounds obtained at a 600°C pyrolysis, and found mainly pyrrole structures. Despite the strong interest of this primary study on tholins pyrolysis, coupled with a powerful GC-GC-MS analysis, the quasi non-detection of products at 250°C remains a major issue. Indeed, the pyrrole structures detected at 600°C are probably not representative of the original solid material, as it is known that the aromaticity of pyrolysates increases with the temperature [21]. Actually, recent NMR studies on tholins analysis confirmed a low aromatic content of tholins [22-24].

Previous thermal degradation studies of tholins confirmed the need for soft pyrolysis temperatures to address the issue of a tholins molecular identification representative of the chemical structure of the native non-heated material [25-26]. The study by Nna Mvondo et al. [26] pointed out that Titan's tholins contain more open-chain structures than ring-shaped structures and that cyclisation occurs during high temperature treatments. Similarly, He et al. [25] showed that the heated solid residue has an elemental content significantly modified for temperatures higher than 300°C. There is therefore a need for coupling a sensitive organic compound analytical technique to a soft temperature pyrolysis extraction.

This is the purpose of the present work, in which we introduce an alternative analysis method based on PhotoElectron/PhotoIon Coincidence (PEPICO) techniques coupled with tuneable VUV synchrotron radiation, and using a soft pyrolysis extraction method, by heating the material at temperatures lower than 300°C. This experimental method has already been successfully employed for species identification, including isomer differentiation, in complex media [27-30] and is applied here to improve the characterization of the chemical composition of Titan tholins.

**Experimental Setup and Methodology**

**Sample synthesis**

Tholins samples were produced in the PAMPRE radio-frequency capacitively coupled plasma reactor described in detail in Szopa et al. [10]. In this work, the experimental conditions at room temperature were a $N_2$-$CH_4$ gas mixture containing $5.0 \pm 0.1\%$ of methane at a flow rate of $55.0 \pm 0.1$ sccm, a pressure of 0.9 mbar, and a plasma power of 30 W. During the production process, the particles were gently deposited in a glass vessel without any interaction with the substrate and were collected into microvials for further analysis.

**Vaporization and chemical analysis on the VUV DESIRS-synchrotron beamline**

The chemical analysis was performed at the DESIRS undulator beamline (Synchrotron SOLEIL, France) [31], on the SAPHIRS permanent molecular beam endstation, which is composed of two chambers, expansion and ionization, connected through a 0.7 mm skimmer [31].

Tholins samples were heated in an in-vacuum oven placed inside the expansion chamber, and the resulting vapor was mixed with 1 bar of Argon and expanded through a 70 μm nozzle. The oven temperatures were chosen according to tholins thermal stability [25]. Temperature has to be above 150°C to go over the simple extraction of adsorbed water. But the native compounds in the aerosols are increasingly altered with temperature, showing thermic cracking effects. Beyond 300°C, both the elemental analysis and infrared signature of the residual heated sample significantly evolve, showing a drastic thermal evolution of the material [25]. Oven temperatures between 150 and 300°C were therefore chosen in the present study as the best compromise to extract as much representative molecules as possible without modifying significantly the material. In this range, three temperatures 180, 230 and 280°C were used in order to probe a possible change in the volatiles compounds collected according to the heating.

After traversing the skimmer, the vapor entered the ionization chamber of SAPHIRS where sits a double imaging PEPICO (i$^2$PEPICO) spectrometer, named DELICIOUS III [32], which combines a velocity map imaging (VMI) apparatus [33] with a modified ion Wiley-McLaren time of flight analyser/imager. The coincidence scheme is used here to filter the electrons according only to the ion mass and thus to provide mass-selected photoelectron images. Abel inversion of the images [34] yields the corresponding photoelectron spectra (PES) for all the ions in the mass spectra, in a multiplex manner, with an estimated resolution of 200 meV. The error bars on the experimental PES are estimated assuming a Poisson distribution on each independent pixel of the photoelectron images, and subsequent propagation of the error over the algebra operations performed by the Abel inversion algorithm. These error bars are then used by the least squares fit routine to estimate the statistical error associated to the relative isomeric abundance.

The molecular beam was ionized by the VUV synchrotron radiation at the center of the spectrometer. The monochromator was set to deliver 3 x 10$^{12}$ photons/sec with a resolution of 10 meV at 10 eV. A gas filter [35] located upstream the monochromator and filled with Ar ensured spectral purity by effectively absorbing the high harmonics radiation emission from the undulator. In practice, i$^2$PEPICO data were recorded at the 4 fixed photon energies of 9.5 eV, 10.5 eV, 11.5 eV and 12.5 eV for a typical 7200 sec duration.

**Molecular identification**

Molecular identification of selected chemical compounds was achieved through the comparison of the cation electronic signatures in our experimental results with existing experimental data. All $C_xH_yN_zO_w$ species were considered: oxygenated compounds are part of the original tholins aerosols, as oxidation of the surface occurs in the time between production and characterization. The oxygen content in our samples has been previously studied and quantified [21, 36-37].

The mass resolving power of DELICIOUS III under the present experimental conditions (time-focusing mode operation) is estimated from the most intense peak at m/z 111 to be $M/\Delta M = 450$ at the full width half maximum (FWHM). This value is not high enough to directly infer the chemical composition of the molecules.

i. **Photoelectron spectra (PES)**

The isomers identification is made by comparison between mass selected photoelectron spectra obtained in this work from the vaporized tholins and the He I, He II and Ne photoelectron spectra existing in the literature. This identification can be hindered due to the following obstacles: (1) the presence in the spectra of electronic structures corresponding to precursors of species produced by a dissociative photoionization process; (2) internal temperature effects (imperfect cooling) that may lead to hot bands, and in some cases to the presence of several conformers; (3) the lack of defined structure in the PES and the overlap between different isomers; (4) the absence of experimental PES data for the isolated molecules in the literature, and more commonly, the lack of fragment selected data to deduce the state-selected fragmentation pattern; and (5) the increasing number of possibilities with increasing mass.

Clearly the higher the mass, the more difficult it is to attribute the isomers, and therefore this work is dedicated to the isomer attribution of the lighter species of the tholins composition with m/z ≤ 69.

ii. **Orbitrap High Resolution Mass spectrometry analysis (OHR-MS)**

Complementary analyses using OHR-MS were made with a hybrid linear trap/orbitrap mass spectrometer (LTQ orbitrap) of the polar species, following the methodology developed in [12]: a soft electrospray ionization, but limited to the soluble fraction of the sample, coupled to a high resolution mass analyzer. This analysis assumes that the soluble fraction is representative of the bulk tholins (as shown in [17]). The lower mass limit of m/z 50 with this instrument precludes a full comparison with the PEPICO data for light species.

**Results and discussion**

**Effect of the oven temperature on the volatiles compounds extracted from the tholins sample**

**Figure 1** (a-c) shows the mass spectra obtained at 10.5 eV of photon energy with DELICIOUS III for oven temperatures of (a) 180°C, (b) 230°C and (c) 280°C obtained with the same acquisition time. The detected ions are the same in the three mass spectra. The major quantitative difference between these results is the ratio between the heavier and lighter ions, which increases as a function of the temperature. Looking at the absolute scales in Figure 1 this increase correspond to both a gain of the heavier masses (>100 amu), and a loss of the lighter ones. While the temperature trend of the heavier masses is intuitive, the loss of signal on the lighter ones must be explained by the volatile fraction gradually disappearing with time, since the increase in temperature shown in Figs. 1a-1c followed a chronological order. In any case, we did not observe any significant temperature (or time) dependence on the measured PES on the most part of the species analyzed in this work, which leads us to conclude that the evolution of this ratio should be simply associated with the relative vapor pressure and quantity of these species. These effect are most probably attribute to the increase of the temperature, however we note that it could be partially attributed to a time effect, where the volatile compounds would gradually disappear, since the increase of the temperature follows the chronologic order of the data acquisition.

The invariability of the PES with the temperature indicates that the structure of the molecules extracted from the sample at temperatures between 180°C and 280°C is unchanged. Therefore, and in order to increase the data statistics, we added the data recorded at the two lowest oven temperatures, 180°C and 230°C. Even if we have no evidence of a possible degradation of the sample at 280°C, on a precautionary principle, we discarded, for the PES, the analysis of the corresponding data.

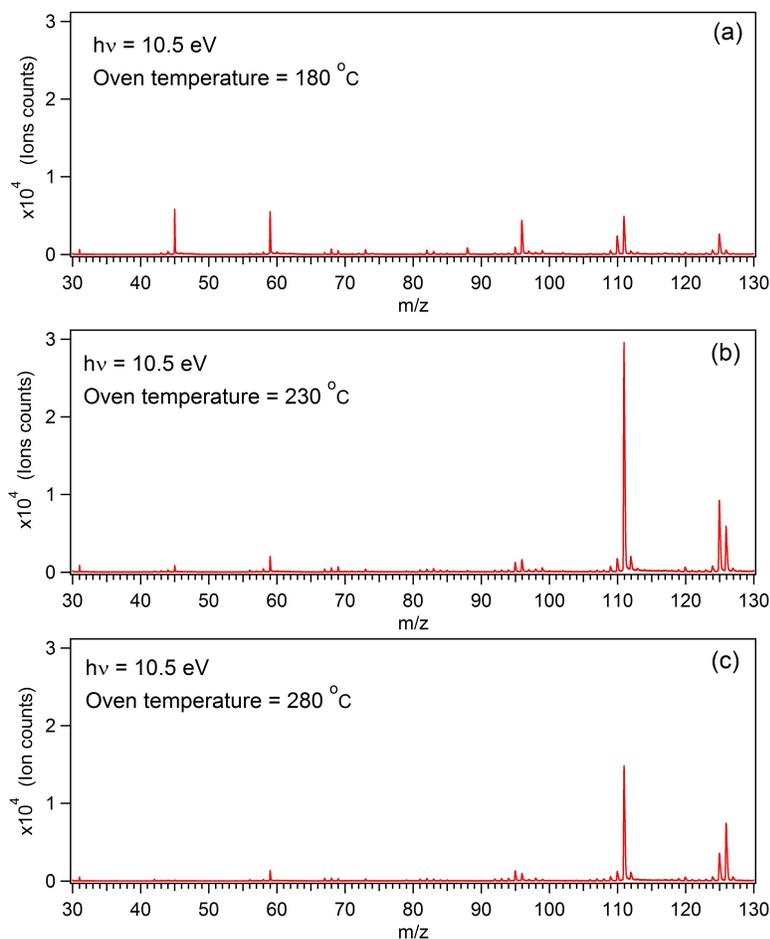

**Figure 1**: Mass spectra obtained during 7200s at 10.5 eV photon energy at DESIRS beamline with an oven temperature of (a) 180°C, (b) 230°C and (c) 280°C.

**General comparison with OHR mass spectra**

**Figure 2** (a) shows a typical OHR mass spectrum of the soluble part of tholins aerosol. The OHR mass spectrum presents two peak groups: (1) strong single peaks are visible for m/z < 130, (2) a polymeric structure visible with regular patterns for m/z > 130. Figure 2 (b) is a VUV mass spectra (VUV-MS) obtained at the DESIRS beamline for volatile molecules released from tholins after soft heating at 180°C. Detected ions are globally consistent between the two spectra, showing no major differences according to the extraction method except for quite different relative intensity ratio. The polymeric structure with regular patterns for m/z > 130 observed during the soluble fraction analysis was also detected in the VUV-MS (**Figure 2b**) but was not analyzed due to the low signal intensity.

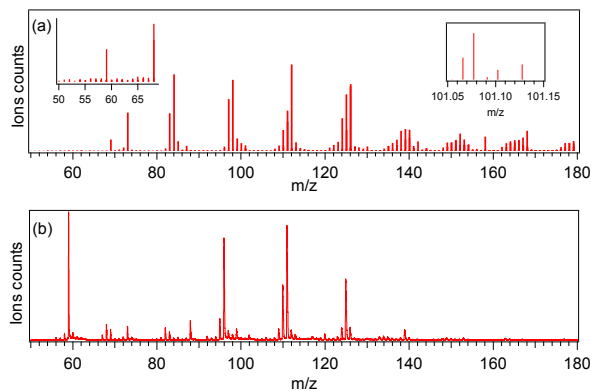

**Figure 2:** (a) OHR mass spectrum (b) VUV Mass spectra (VUV-MS) obtained at 10.5 eV photon energy with an oven temperature of 180°C on the DESIRS beamline with DELICIOUS III spectrometer.

**Molecular identification of the lighter ions, with m/z < 69**

Figure 3 corresponds to the VUV-MS for the signal of the lighter ions at m/z ≤ 69 obtained at 10.5 eV photon energy with an oven temperature of 230°C. Several ions are detected, such as m/z 17, 30, 31, 42, 43, 44, 45, 56, 57, 58, 59, 60, 67, 68 and 69. The molecular identification of some of these ions—those where the statistics are high enough to extract the PES—is performed by comparing the PES obtained in coincidence with PES reference spectra, as detailed in the experimental section.

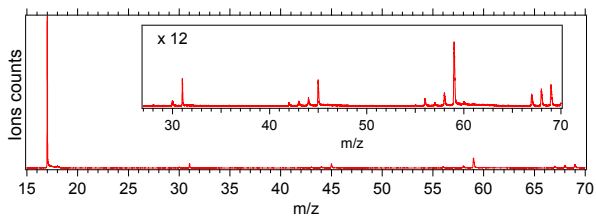

**Figure 3**: VUV-MS obtained at 10.5 eV at an oven temperature of 230°C. The inset shows a close-up of the region between m/z 27 and 70.

First, the ion at *m/z* **17** is found unambiguously to be ammonia (NH$_3$) by comparing our photoelectron spectra with the one obtained by Locht et al. [38]. Thanks to this attribution we could use 8 observed vibrational transitions to calibrate the kinetic energy scale of our spectra, leading to an absolute precision of 5 meV on the binding energy for all species.

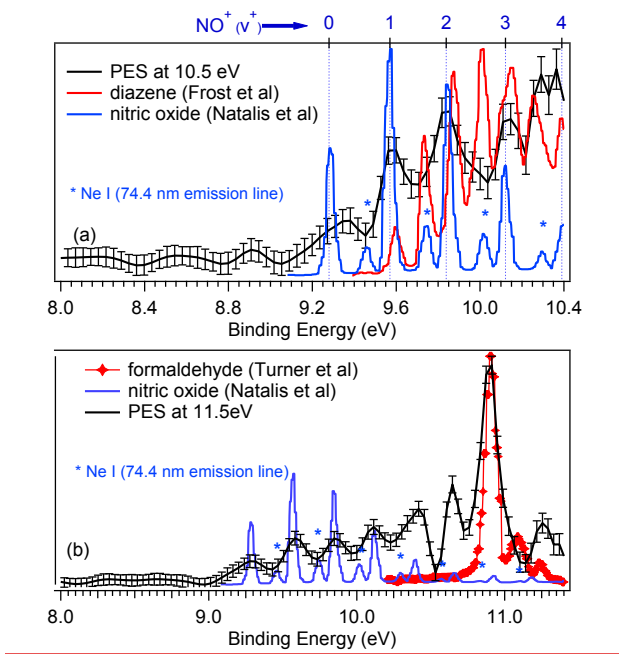

**Figure 4** : Photoelectron spectra for ion at m/z 30 obtained in this work (black line) at (a) hν = 10.5 eV and (b) hν = 11.5 eV, along with those for the possible isomers found in the literature for the formaldehyde (CH$_2$O) (resolution 80 meV)) [39], diazene (N$_2$H$_2$) [40] represented by red line and nitric oxide (NO) (resolution 15 meV) [41] ionization energies represented in blue lines.

In the case of ion at *m/z 30* several species with this m/z ratio could make a contribution to the observed PES. Comparison of the PES (**Figure 4**) found in the literature with our spectra suggests the presence of nitric oxide (NO) [41], and formaldehyde (CH$_2$O) [42], the diazene (N$_2$H$_2$) [40] photoelectron spectrum is also represented in **Figure 4** (b), but the comparison suggests that if the diazene is present it is in small quantity, since our spectra are dominated by NO structures.

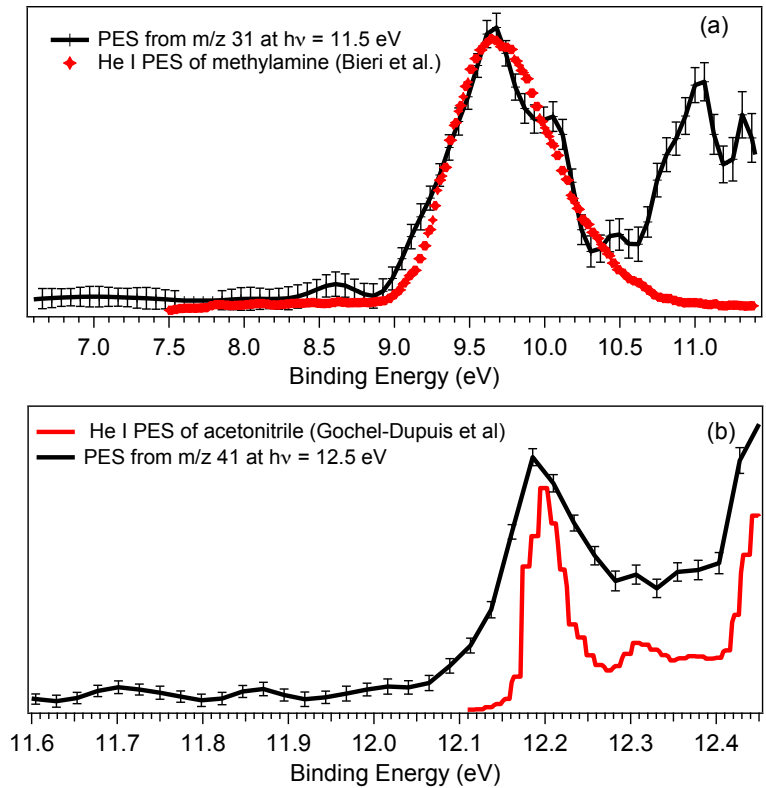

**Figure 5**: (a) Photoelectron spectra obtained in this work (black line) for m/z 31 at 11.5 eV photon energy and He I photoelectron spectra (red marker) obtained of methylamine (CH$_5$N) by Bieri et al. [43]. (b) Photoelectron spectra obtained for m/z 41 at 11.5 eV photon energy and He I photoelectron spectra obtained of acetonitrile (CH$_3$CN) by Gochel-Dupuis et al (resolution 25 meV) (red line) [44].

A more complex example of the isomers identification in this work is illustrated in **Figure 5** for the ion at *m/z* **31**. Although we could readily identify the methylamine (CH$_5$N) and probably nitrosyl hydride (HNO) through comparison with the experimental PES by Bieri et al. [43] and Baker et al. [45], the region above 10.8 eV is unexplained. Other candidate such CH$_3$O radical [46-47] was also excluded by their cation's electronic footprint, and the fact that radicals are not expected in tholins. Alternatively, the structure above 10.8 eV could come from dissociative ionization of some unidentified heavier ion(s).

For *m/z* **41** we could easily identify the contribution as acetonitrile (CH$_3$CN), as illustrated on the comparison between our PES obtained at 12.5 eV and the He I PES of the acetonitrile obtained by Gochel-Dupuis et al. [44] on **Figure 5** (b). For the weak ion signal at *m/z* **42** (**Figure 6**) we identify the possible presence of 3H-diazirine (cyclic CH$_2$N$_2$) [48], Cyanamide (NC-NH$_2$) [49] and propene (C$_3$H$_6$) [50] and exclude the presence of cyclopropane (C$_3$H$_6$) [51] and diazomethane (linear CH$_2$N$_2$) [52]. N$_3$ radical [53] is unlikely, but cannot be excluded, For binding energies greater than 11 eV we observed structures of unidentified origin.

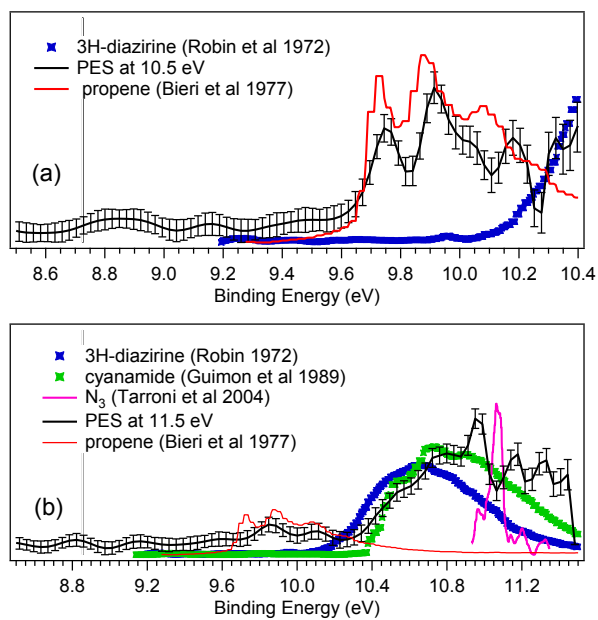

**Figure 6** : Photoelectron spectrum for ion at m/z 42 obtained in this work at (a) hν = 10.5 eV and (b) hν = 11.5 eV, along with those for the possible isomers found in the literature for 3H-diazirine (CH$_2$N$_2$) (resolution 30 meV) (blue marker) (47), propene (C$_3$H$_6$) (red line) (48), N$_3$ (pink line) [53] and cyanamide (CN$_2$H$_2$) (green marker) [49].

For the ion at *m/z* **43** we could compare the PES obtained in this work with three isomers of C$_2$H$_5$N, the N-methylmethylenimine, C-methylmethylenimine and ethylenimine. Our experimental PES obtained at 10.5 eV could be explained by the presence of only C-methylmethylenimine isomer as illustrated on the PES on **Figure 7**. However, we have to take into account two important points: (1) First one, despite the C- methylmethylenimine spectra was obtained by Bock et al [54] with a higher resolution (18meV-25meV) than the one obtained in this work, we verify that we have more resolved electronic structures. In our point of view, we interpret this discrepancy as a higher temperature effect on the C-methylmethylenimine spectra due to their synthesis procedures that involves pyrolysis process without an adiabatic expansion afterwards [54]. (2) Second one, since this molecule is highly unstable, [55] showing a tendency to polymerize, we believe that if it is present it would be a product of thermal decomposition of some unspecified compound in the tholins.

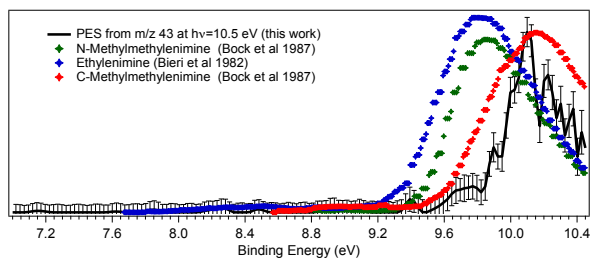

**Figure 7**: Photoelectron spectrum for ion at m/z 43 obtained in this work at hν = 10.5 eV along with those for the possible isomers found in the literature for $C_2H_5N$, the N-methylmethylenimine (green markers), C-methylmethylenimine (red markers) [54] and ethylenimine [43] (blue markers).

The analysis of our PES spectrum obtained for the ion at m/z 44 indicates the possible presence of acetaldehyde ($CH_3CHO$), propane ($CH_3CH_2CH_3$) and ethylene oxide ($C_2H_4O$), as illustrated in red, green and blue, respectively on the PES of **Figure 8**. However, this three species do not describe the whole PES spectrum obtained in this work, for example for the structures observed around 11 eV binding energy, so other unidentified species contribute to this channel. The presence of vinyl alcohol ($CH_2CHOH$) [56], methyl diazene ($CH_4N_2$) [40] and the three $C_2H_6N$ isomers dimethyl amidogen [57], $CH_3CHNH_2$ [58] and $CH_2NHCH_3$ [58] can be discarded due to their lower ionization energy.

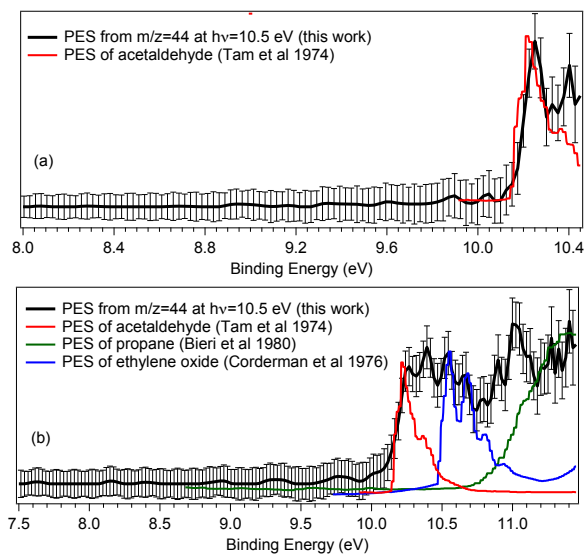

**Figure 8** : Photoelectron spectrum for ion at m/z 44 obtained in this work at (a) hν = 10.5 eV and (b) hν = 11.5 eV (black line) and photoelectron spectrum for acetaldehyde (CH$_3$CHO) obtained by Tam et al (red line) [59], for propane (CH$_3$CH$_2$CH$_3$) by Bieri et al (resolution between 60 and 100 meV) (green line) [60] and for ethylene oxide (C$_2$H$_4$O) by Corderman et al (resolution 25 meV) (blue line) [61].

**Figure 9** (a) and (b) present the isomers identification conducted for the ion at *m/z* **45**. Despite the difference in resolution and in the photon energy between our spectra and the He I and He II ones, the very satisfactory matching between the cation's electronic footprint of ethylamine (C$_2$H$_7$N) and formamide (HCONH$_2$), and our recorded spectra shows unambiguously that these two molecules are present in the vaporized fraction of the tholins, with formamide being the most abundant by far.

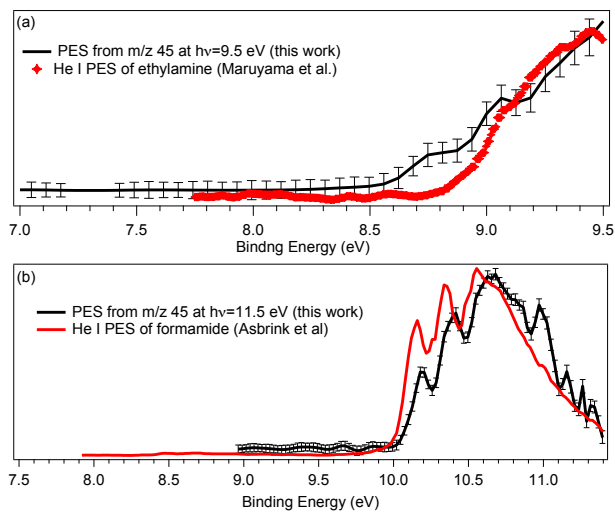

**Figure 9**: Photoelectron spectra obtained in this work (black line) and He I and He II photoelectron spectra (red markers and red line) found in the literature. The black lines correspond to the photoelectron spectra for ion at m/z 45 obtained from vaporized tholins at 9.5 eV (a) and 10.5 eV (b) photon energies while the red markers represent (a) the He I (21.21 eV) photoelectron spectra obtained for the ethylamine ($C_2H_7N$) (resolution 60 meV) by Maruyama et al [62] and (b) the red line for formamide ($HCONH_2$) by Asbrink et al. [63].

For medium-sized ions that can be studied with the OHR-MS technique (m/z > 50) but that are small enough to have a manageable number of possible isomers, it is interesting to compare the information obtained with the OHR-MS and VUV-MS analysis.

For instance, for the ion at *m/z 56*, the aminoacetonitrile (AAN) is the only candidate offered by the OHR-MS studies [12]. Several isomers with known electronic footprints have been considered here to model the PES associated with this mass (**Figure 10**), including AAN. However, although the subsequent least squares fit predicts the presence of some amount of 2-propenal ($C_3H_4O$) and AAN, it is clear that most of the obtained PES is not well fitted or explained, including the region where AAN would contribute, and the binding energies above 11 eV. Expanding further on the presence of AAN, Bellini et al. (55) recently recorded the fragmentation pattern of AAN$^+$ and found that, at the photon energy of 11.5 eV, the main fragment ion would be m/z 29 (around 25%), which it is not in agreement with our VUV-MS recorded at this energy, where we found a branching ratio of around 11%. Assuming that ion at m/z 29 comes exclusively from dissociative ionisation of AAN places an upper limit of 44% to the contribution of AAN to the *m/z* 56 channel. Of course, *m/z* 29 could come from dissociation of other parent ions (although not from HCO or C$_2$H$_5$ since the presence of radicals is ruled out), and indeed the least squares fitting provides a AAN contribution well below 44 %. We note that AAN is highly thermolabile as reported by Bellini et al. (55), and decomposes at room temperature in a matter of hours so that not all of the AAN might survive the vaporization temperatures used in this work, while the softer electrospray method used in the OHR-MS is more favorable for fragile molecules, provided they are soluble. We cannot explain, however, why McGuigan et al. [18] using a high temperature pyrolysis found AAN to be a major species, while in our work it represents a minor fraction of the m/z 56 channel. Other structures like 2-propenal (C$_3$H$_4$O) and cyclobutane (C$_4$H$_8$) are likely, but a large portion of the PES remains unexplained so that other isomers than the ones named in **Figure 10** must be present.

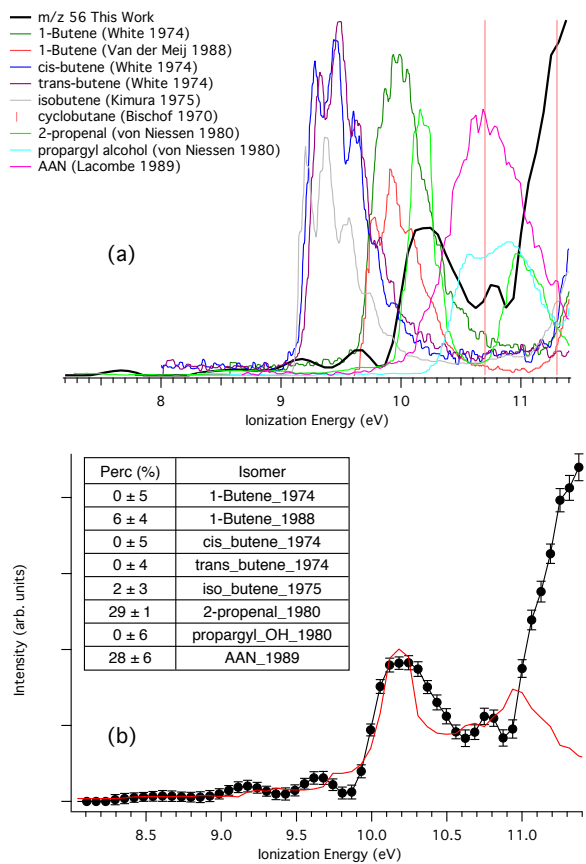

**Figure 10**: (a) Photoelectron spectrum for ion at m/z 56 obtained in this work at hυ = 11.5 eV, along with those for the possible isomers found in the literature. [64-69]. The curves have been normalized to their integral over the displayed ionization energy range. (b) Experimental PES (black circles) and least squares fit of the linear combination of all isomers shown in panel (a) to the data. The contributions of each isomer to the data are shown in the inset.

In the same manner, ions at *m/z 58* and *59* have been assigned by OHR-MS to acetamidine ($C_2H_6N_2$) and guanidine ($CH_5N_3$), respectively. This cannot be fully verified in our PEPICO data since no information whatsoever was found on the photoionization of acetamidine, and only ionization energies exist for guanidine. From **Figure 11**, we find, however, that several other isomers can contribute to the PES of m/z 58, without any species being predominant. Indeed, we succeeded in obtaining a quite satisfactory fitting of the PES by a by a linear combination of all possible isomers. The departure from a perfect fitting, indicates that other molecules apart from the ones tried here may contribute to the signal. In order to test for the presence of acetamidine, we have calculated its adiabatic ionization energy at the PBE0/aug-cc-pvdz DFT level, and obtained a value of 8.47 eV. The geometry change upon ionization is slight (less than 5 % for chemical bands and less than 6 % for angles), so one would expect a marked adiabatic transition. As seen in **Figure 11**, this value is far from the main experimental band starting at 9.4 eV, but could be in agreement with the weak structure observed around 8.5 eV. Although more information than just one theoretical point is needed to corroborate its presence, we can nevertheless conclude that if acetamidine is present, it is only in very small concentrations. Since this molecule is predicted in the OHR-MS, the disagreement could be attributed to thermal decomposition. Note that the measured vapor pressures at 25°C by the ACD (176 torr) and EPISuite (22 torr) are relatively high so that, barring decomposition, we would expect a strong signal at our working temperatures. Indeed, acetamidine has been reported to decompose at 95°C to yield 70% of acetonitrile [70], which is consistent with both the fact that acetamidine is not present (or in trace quantities) in our measurements, and that acetonitrile is detected in the measurements performed at 12.5 eV (see **Figure 5b**).

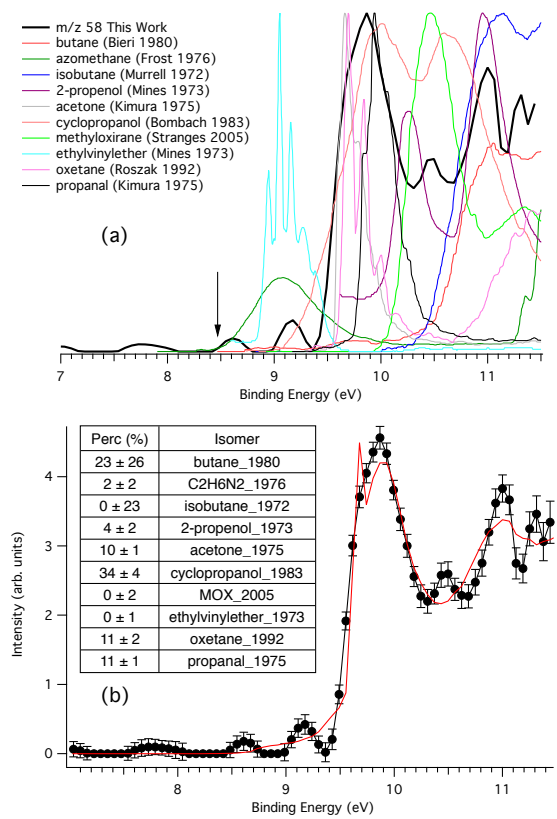

**Figure 11**: (a) Photoelectron spectrum for m/z 58 obtained in this work at hυ = 11.5 eV, along with those for the possible isomers found in the literature. [40, 60, 65, 71-75] The curves have been normalized to their integral over the displayed ionization energy range. The calculated adiabatic ionization of acetamidine (see text for details) has been marked with an arrow. (b) Experimental PES (black circles) and least squares fit of the linear combination of all isomers shown in panel (a) to the data. The contributions of each isomer to the data are shown in the inset.

For the ion at *m/z 59* (PES shown in **Figure 12**), which dominates our VUV-MS (low mass side) at all temperatures (see **Figure 1**), a good fit of the PES could be achieved with the available isomers, and acetamide ($C_2H_5NO$) is found as the most abundant contribution. Although the presence of acetamide is in agreement with the OHR-MS data, it was found to be minor compared with the main species at m/z 59, guanidine ($CH_5N_3$). However, guanidine is totally absent from the PES spectra. The experimental electron impact ionization energy of guanidine is 9.10 eV [76], and its recently calculated adiabatic and vertical first ionization energies are 8.48 and 9.09 eV respectively [77]. As in the case of AAN, the reason for this discrepancy is found in the extraction methods. Guanidine is a fragile molecule which decomposes at temperatures above 160°C [78]. The present thermal extraction method is therefore inappropriate for the detection of this specific molecule.

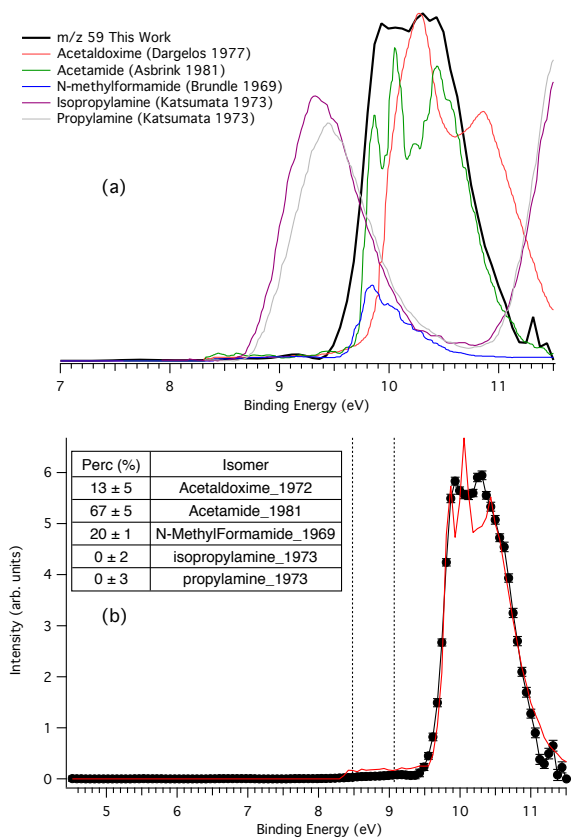

**Figure 12**: (a) Photoelectron spectrum for ion at m/z 59 obtained in this work at hv = 11.5 eV, along with those for the possible isomers found in the literature [63, 79-81]. The curves have been normalized to their integral over the displayed ionization energy range. (b) Experimental PES (black circles) and least squares fit of the linear combination of all isomers shown in panel (a) to the data. The contributions of each isomer to the data are shown in the inset. The dashed lines represent the range of energies where signal from guanidine should be seen according to literature [76-77].

**Figure 13** shows the PES corresponding to ions at *m/z 69*. The lack of sharp structure in our experimental PES—a result of the limited experimental resolution and/or the congestion due to the presence of different molecules and/or conformers and/or vibrational congestion—and the number of possible isomers make the identification challenging. It is nevertheless clear that all of the experimental PES can be modeled with the 1H-1,2,4-triazole isomer, which is in agreement with the OHR-MS experiments that establish the molecular composition as $C_2H_3N_3$. The VUV-MS data rule out the presence of another possible isomer, the 1H-1,2,3-triazole.

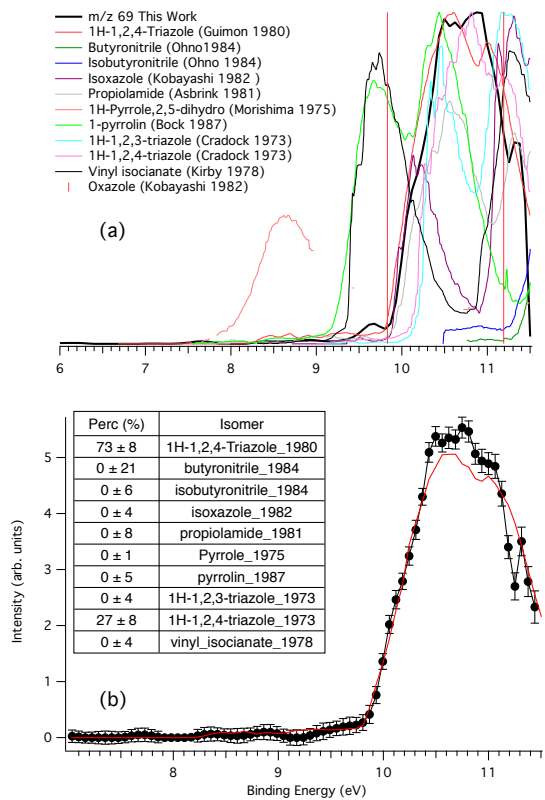

**Figure 13**: (a) Photoelectron spectrum for ion at m/z 69 obtained in this work at hv = 11.5 eV, along with those for the possible isomers found in the literature [54, 63, 82-87]. The curves have been normalized to their integral over the displayed ionization energy range. (b) Experimental PES (black circles) and least squares fit of the linear combination of all isomers shown in panel (a) to the data. The contributions of each isomer to the data are shown in the inset.

**Discussion**

The Aerosol Collector and Pyrolyser-Huygens instrument detected ammonia ($NH_3$, m/z 17) and hydrogen cyanide (HCN, m/z 27) as the main released molecules by pyrolysis at 600°C of the aerosols in the atmosphere of Titan. In Coll et al [88] it was shown that Titan's tholins produced in "cold plasma" conditions, such as in the PAMPRE reactor, provided the best possible analogues to reproduce the volatiles released with the ACP experiment. Our work, focused on the light species extracted from PAMPRE tholins through soft pyrolysis conditions (T<240°C), provides further insights on this in-situ thermal extraction of light molecules and on the possible presence of larger structures in Titan's aerosols.

The ions detected at m/z ≤ 69 in this work are summarized in Table 1. Eleven ion signatures with one to five heavy atoms were detected. Most of them could be firmly identified with their corresponding PES. A few oxygen-bearing compounds, NO, $CH_2O$ (formaldehyde), $C_2H_4O$ (acetaldehyde and ethylene oxide), $HCONH_2$ (formamide) and $CH_3CONH_2$, are reported, confirming the oxygen content of the samples [37]. The O-bearing molecules are always found in these kinds of laboratory analogues [21]. Pyrolysis process is discarded as the oven is under vacuum. A recent chemical analysis of the samples by XPS highlights that the O-bearing molecules are only present in the external layer of the solid sample, not in its core [89]. They are produced when the sample is exposed to ambient air. High resolution mass spectrometry shows that the O-bearing molecules have a similar mass pattern to the rest of the sample [12]. Those are therefore no atmospheric volatile organic compounds adsorbed on the sample during the transfer, but result from an oxidation process. As they are considered as a contamination, we did not discuss further their formation in the manuscript.

The strong signature of ammonia in our analysis (**Figure 3**) is in agreement with the ACP data. Ammonia is a volatile molecule, which possibly results from the pyrolysis of primary terminal amine functional group in the macromolecules composing tholins. Such an origin is consistent with the identification of light primary amines in our work. We detected the two simplest primary amines, methylamine and ethylamine at m/z 31 and 45, but also the amino-acetonitrile at m/z 56. It is well known that the amine functional group plays a key role in prebiotic chemistry. The amino-acetonitrile molecule has moreover its amine-functional group in alpha-position regarding the nitrile group. The hydrolysis of this precursor leads to the formation of glycine, the smallest α-amino-acid. Amine functional groups can require specific targeted analytical techniques [90] and the present analysis appears to be efficient for their detection and suggests an amine-origin of the $NH_3$ signature detected after pyrolysis of Titan's aerosols by the ACP instrument.

The second main molecule detected by ACP, hydrogen cyanide, could not be identified here because its ionization energy is higher than the photon energy used for this work. Tholins are known to bear nitrile functional groups through their mid-IR signature around ~2200 cm-1 [3, 91]. We have identified in this work at m/z 41 the presence of acetonitrile ($CH_3CN$). A similar pyrolysis extraction from tholins was performed in [25] with a simple QMS analysis of the released molecules. No firm identification was possible in this case, but the m/z 27 ion signature could reasonably be attributed to HCN. The evolution of its ion intensity was monitored according to the increasing temperature, showing a progressive release of this molecule, actually negligible at soft temperatures (T<300°C).

| m/z | Soft-Pyr-MS-PES | Solvent-OHR-MS |
|---|---|---|
| 17 | $NH_3$ | - |
| 30 | NO | |
|  | $CH_2O$ (formaldehyde) | |
| 31 | $CH_3NH_2$ (methylamine) | - |
| 41 | CH$_3$CN (acetonitrile) | |
| 42 | Cyclic CH$_2$N$_2$ (3H-diazirine) | - |
| | CN$_2$H$_2$ (cyanamide) | |
| | C$_3$H$_6$ (propene) | |
| | N$_3$ (?) | |
| 43 | C$_2$H$_5$N (C-methylmethylenimine) | - |
| 44 | CH$_3$H$_2$CH$_3$ (propane) | - |
| | C$_2$H$_4$O (acetaldehyde and ethylene oxide) | |
| | + Others | |
| 45 | C$_2$H$_5$NH$_2$ (ethylamine) | - |
| | HCONH$_2$ (formamide) | |
| 56 | NH$_2$CH$_2$CN (aminoacetonitrile) | C$_2$N$_2$H$_4$ |
| | + Others | |
| 58 | Mixture | C$_2$H$_6$N$_2$ |
| 59 | CH$_3$CONH$_2$ (acetamide) | CH$_5$N$_3$ |
| | CHONHCH$_3$ | |
| 69 | 1H-1,2,4-triazole | C$_2$H$_3$N$_3$ |

**Table 1**: Volatile molecules with m/z ≤ 69 extracted from PAMPRE Titan's tholins after a soft heating at ~230°C and identified by PES spectroscopy in this work (Soft-Pyr-MS-PES). Comparison with the elemental formula of the main species detected by OHR-MS at the same mass units, when possible (m/z > 50), and extracted from tholins in methanol.

The molecules identified in this work are compared with the elemental formula of the main species detected by OHR-MS at the same mass unit for m/z > 50 in Table 1. First we notice that the species are mostly consistent in spite of the different extraction method, in agreement with the previous general comparison of the mass spectra (**Figure 2**).

The main difference is for m/z 59, corresponding to acetamide in the case of the soft-pyrolysis-MS-PES analysis and to CH$_5$N$_3$ in the solvent-OHR-MS analysis, strongly suspected to be guanidine. Guanidine is a thermally fragile molecule that decomposes with pyrolysis-extraction methods even with the mild temperatures used here. This underlines the complementarity of OHR-MS and the resistive heating described here. Indeed species that are not seen in the former due to the lack of solubility can be observed by resistive heating combined with PES, and conversely, thermally fragile molecules that are not seen at high temperatures, are only seen in OHR-MS provided that they are soluble.

For the other species, the elemental compositions obtained with OHR-MS are consistent with the structure extracted from the MS-PES. We confirm the putative attribution made for m/z 56 in [12] as amino-acetonitrile. Moreover we highlight a N-aromatic structure for the species at m/z 69 with a triazole ring. This cycle involves two adjacent double-bonded nitrogen atoms, which are also seen in diazirine, a smaller structure identified at m/z 42. Tetrazolo[1,5-b]pyridazine, a gas product with similar chemical properties was moreover previously detected in the plasma discharge where tholins are produced [92]. This specific N=N pattern is possibly a marker of a direct incorporation of $N_2$ (or $N_2^+$) in the growing organic material composing tholins. This basic structure would deserve further investigation to interpret the N-aromatic signature globally detected in Titan's tholins [22, 91] and to understand the role of nitrogen in the organic growth in the atmosphere of Titan. The present knowledge on gas-phase nitrogen chemistry representative of Titan's atmosphere actually hardly explains the incorporation of two adjacent nitrogen atoms in the organic products [93].

**Conclusion and perspectives**

The chemical composition of the volatile fraction of Titan's tholins was studied by means of electron/ion coincidence techniques coupled to tunable VUV synchrotron radiation. A large number of ions were detected in the mass spectra and for each of them a photoelectron spectrum was obtained through the coincidence scheme. For some of the lighter ions ($m/z \leq 69$), comparison with existing experimental PES yielded the structure, while for heavier ions no analysis was performed due to the lack of PES experimental data and the larger number of possibilities. This limitation highlights the importance of recording experimental data on the photoionization of isolated relevant molecules, such as electron spectroscopy, state-selected photodissociation and total ionization cross section for quantitative analysis. In addition, although thermolysis data could provide further speciation and also be invoked to explain the differences between HR-OMS and the resistive oven PEPICO experiments, the lack of literature on the subject and the expected overlap of the thermolysis products, especially at low masses, prevent us from attempting so.

Crossed analyses, addressing complementary fractions of the material have to be used to fully understand the complex composition of tholins. The analytical methodology developed in this work was focused on relatively mild conditions in order to ensure that the molecules extracted and softly ionized are representative of the bulk material: a temperature lower than 230°C and ionizing energies lower than 12.5 eV.

Beyond the present case of tholins analysis, the present methodology consisting in coupling a soft pyrolysis to an imaging PEPICO set-up associated to VUV synchrotron radiation could be extended to other solid materials, including organic powders, in a very complementary Way with other analytical techniques.

AUTHOR INFORMATION

**Corresponding Author**

Electronic mail: nathalie.carrasco@latmos.ipsl.fr

**Present Addresses**

†Present address: Laboratoire de Chimie Physique, Matière et Rayonnement, Université Pierre et Marie Curie, 11 rue Pierre et Marie Curie, F-75231 Paris Cedex 05

**Author Contributions**

The manuscript was written through contributions of all authors. All authors have given approval to the final version of the manuscript. NC supervised the project. BCM, GG, FGL, BF, AM, LN and NC contributed to the experimental campaigns on the synchrotron. TG and PP contributed to the HRMS measurements and comparisons.

**Acknowledgments**

We would like to thank the general technical staff of SOLEIL for running the facility under project n° 20110728 and 20120953. The authors wish to thank Isabelle Schmitz-Afonso and David Touboul from Institut de Chimie des Substances Naturelles (ICSN) for the OHR-MS measurements. B.F. PhD grant is supported by Région Ile-de-France (DIM-ACAV program). NC acknowledges the European Research Council for their financial support (ERC Starting Grant PRIMCHEM, grant agreement n°636829).

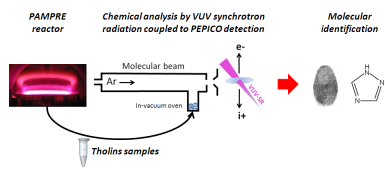

TOC graphic